\documentclass[notoc]{JHEP3}

\title{Vector Meson Photoproduction
from the BFKL Equation I: Theory}

\author{R.\ Enberg$\,^a$, J.R.\ Forshaw$\,^b$, L.\ Motyka$\,^c$ and G.\ Poludniowski$\,^b$ 
\\
$^a$ High Energy Physics, Uppsala University, Box 535, SE-751 21 Uppsala, Sweden
\\ 
$^b$ Department of Physics \& Astronomy, University of Manchester,\\ 
\phantom{$^b$} Manchester M13 9PL, UK
\\ 
$^c$ Institute of Physics, Jagellonian University, Reymonta 4, 30-059 Krak\'ow, Poland}

\abstract{Diffractive vector meson photoproduction accompanied by proton
dissociation is studied for large momentum transfer.
The process is described by the non-forward BFKL equation, for which
a complete analytical solution is found. The scattering
amplitudes for all combinations of helicity are presented. \\
}
\preprint{TSL/ISV-2003-0269}

\usepackage{amsmath}
\usepackage{amssymb}
\usepackage{epsfig}

\newcommand{\be}{\begin{equation}}
\newcommand{\ee}{\end{equation}}

\newcommand{\f}{\ensuremath{\,_2F_1}}

\newcommand{\ba}{\begin{eqnarray}}
\newcommand{\ea}{\end{eqnarray}}

\def\gV{\gamma(\lambda) V(\lambda')}

\def\repart{\mathcal{R}\mathrm{e}\,}
\def\mri{\text}
\def\mq{m}

\renewcommand{\vec}[1]{\mbox{\boldmath $#1$}}
\newcommand{\evec}[1]{\mbox{\scriptsize\boldmath $#1$}}

\newcommand{\beq}[1]{\begin{equation}\label{#1}}
\newcommand{\eeq}{\end{equation}}
\newcommand{\beqar}[1]{\begin{eqnarray}\label{#1}}
\newcommand{\eeqar}{\end{eqnarray}}

\begin{document}

\section{Introduction}

Diffractive vector meson photoproduction at large
momentum transfer, $\gamma p \to V X$, is an interesting process experimentally, because of the clean signal:\ an isolated vector meson with large transverse momentum, that is separated from the proton remnant by a large rapidity gap. Theoretically, it is advantageous because the large momentum transfer $|t|$ provides the hard scale necessary for a perturbative QCD description of the process in terms of hard colour singlet exchange. It has therefore been proposed as an ideal testing ground for BFKL dynamics~\cite{FR,BFLW}.

This process was recently investigated at
HERA~\cite{ZEUS1,Aktas:2003zi}. The ZEUS analysis includes 
the angular distribution of the decay
products, giving access to helicity amplitudes of the
photon--meson transition. 
The measurements gave somewhat
surprising results. Both for $\rho$ and $\phi$, the
dependence of the cross-section on the momentum transfer $t$
is approximately power-like, $|t|^{-n}$, with the exponent
$n \simeq 3$. It can be deduced from the data that the leading 
contribution comes either from the process with no helicity 
flip or from the one with double helicity flip. 
The subleading helicity amplitudes were shown to be
one order of magnitude smaller than the dominant one.

An analysis based on leading-order perturbative QCD predicts
a steeper decrease of the cross-section, $n \simeq 4$, for
the no-flip and double-flip amplitudes~\cite{IKSS}. The
single-flip amplitude $M_{+0}$ gives a differential cross
section with $n \simeq 3$, and should become dominant at
very large momentum transfer $|t|$. It is clear that these
properties are in contradiction with the experimental
results.

Improved theoretical
understanding came about after the
suggestion that the real photon may couple to a
quark--antiquark pair with a significant chiral odd
component~\cite{IKSS}. In general, the quark--antiquark
pair can couple with either chiral even or chiral odd
components, the latter vanishing in perturbation theory in
the limit of massless
quarks. However, it is not clear that the mass should
be interpreted as the current quark
mass. The answer to this question turns out to be
essential for the phenomenology.
Estimates based on QCD sum rules were
performed in \cite{IKSS}, giving a characteristic value of
the non-perturbative mass parameter rather close to the
constituent quark mass for light vector mesons. 
At moderate $|t| \sim 10$~GeV$^2$, the chiral odd amplitude with 
no helicity flip ($M^{\mri{odd}}_{++}$) is then expected to 
dominate. In this paper we compute the chiral odd and chiral
even amplitudes, leaving our results explicitly in terms of 
the quark mass $m$. 
An alternative explanation for the 
dominance of $M_{++}$ and the emergence of dimensional
scaling can be found in \cite{Hoyer}. We shall return to
discuss the connection of our approach with that of
\cite{Hoyer} in our second paper, where we shall focus
on the phenomenology \cite{inprep}.

The analysis of~\cite{IKSS} was restricted to the
lowest order approximation for colour singlet exchange,
that is the exchange of two gluons. In fact, the diffractive
production of vector mesons occurs
at energies $\sqrt{s}$ much larger than the momentum transfer.
In such kinematics, the perturbative QCD corrections
to two gluon exchange are enhanced by powers of large logarithms of
the energy,  $\log(s / |t|)$.
The leading logarithmic corrections are proportional to
$[\alpha_s \log(s/|t|) ]^n$ at the $n$th order of the perturbative
expansion. Thus, it is not sufficient to consider
the lowest order approximation.
A convenient way to perform the resummation of leading
logarithmic terms to all orders in $\alpha_s$  is given
by the Balitsky--Fadin--Kuraev--Lipatov (BFKL) equation \cite{BFKL}.
This approach turned out to be successful in the case
of diffractive production of charmonia at high $|t|$
\cite{FR,BFLW,PSIPSI,FP,HVM}.
Interestingly enough, the $t$-dependence of 
vector meson production was successfully described in \cite{FP}
by a BFKL fit with a non-relativistic wave function
for both light and heavy mesons, we shall comment further on this
finding later.

The purpose of this paper is to present analytic results for the
chiral-even and chiral-odd helicity
amplitudes with complete leading logarithmic BFKL resummation.
This calculation will give control of important QCD
effects in vector meson photoproduction at high energies.
A detailed comparison to available data will be performed in a 
forthcoming paper~\cite{inprep}.

The paper is constructed as follows:
The basic ideas are introduced in Section~2,  
the helicity amplitudes at the lowest order are given in Section~3 and 
their BFKL evolution is found in Section~4. 
Brief conclusions can be found in Section~5.


\section{Hard colour singlet exchange}\label{hcs}

Let us consider a frame in which the photon-proton
collision occurs along the $z$~axis. The momentum transfer
vector $q$ is dominated by its transverse part $\vec q$,
$t = q^2 \simeq -\vec q^2$.
The possible helicity states of the incoming quasi-real photon are
characterized by transverse polarization vectors
\be
\vec \epsilon^{\pm} = {\mp} {1 \over \sqrt{2}}(1,\pm i),
\ee
and similarly for the transversely polarized vector mesons.
For the meson, though, the longitudinal polarization
$\epsilon^{0}$ is also allowed. Thus, the possible photon-meson
transitions may be described using three independent helicity
amplitudes $M_{++}$ (no-flip), $M_{+0}$ (single-flip) and
$M_{+-}$ (double-flip)\footnote{There are also the corresponding amplitudes $M_{--}$, $M_{-0}$ and
$M_{-+}$ which satisfy $M_{++}=M_{--}$, $M_{+-}=M_{-+}$ and $M_{+0}=-M_{-0}$.}.

\EPSFIGURE[t]{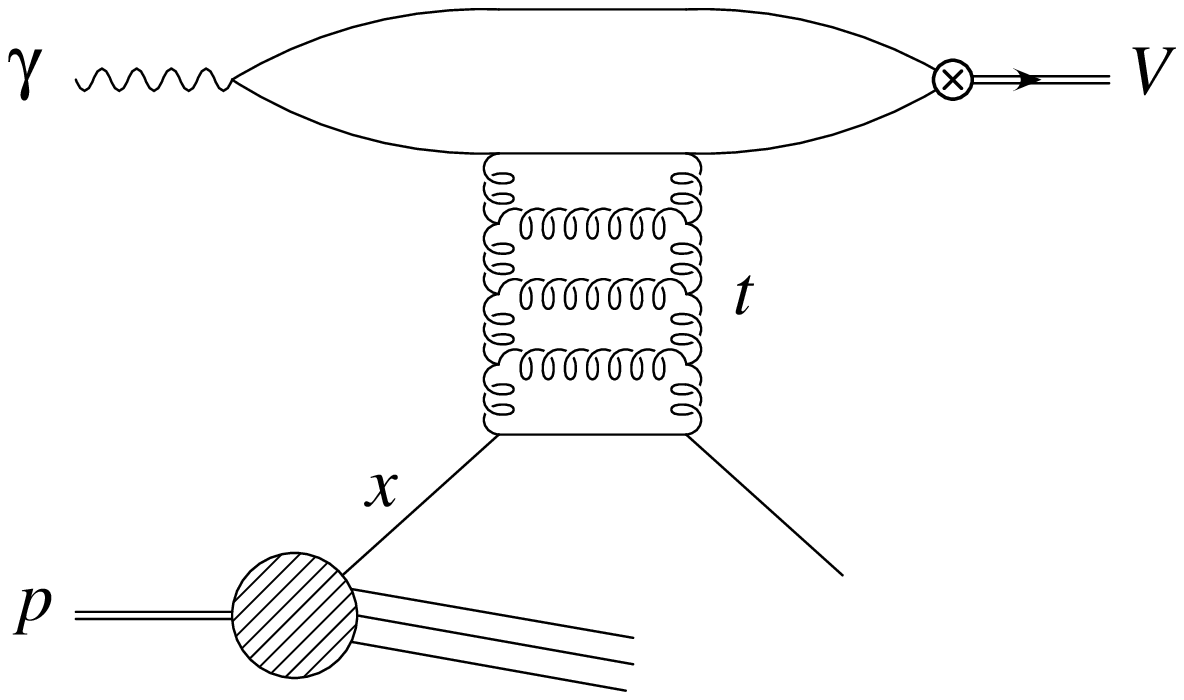,width=0.6\columnwidth}{Vector meson
photoproduction at large momentum transfer via hard colour singlet exchange.\label{fig1}}

The diffractive process $\gamma p \to V X$ at large momentum
transfer $t$ (see Fig.\ \ref{fig1}) takes place by exchange of the BFKL pomeron.
It has been demonstrated that at large momentum transfer, the hard pomeron
couples predominantly to individual partons in the proton \cite{BFLLR}.
Thus, the cross-section may be factorized into a product of the parton
level cross-section and the parton distribution functions,
\begin{align}
\frac{d\sigma (\gamma p \rightarrow VX)}{dt\, dx_j} \;=\;
\biggl(
\frac{4N_{c}^{4}}{(N_{c}^{2}-1)^2}
G(x_j,t)+
\sum_{f}[q_{f}(x_j,t)+\bar{q}_{f}(x_j,t)]\biggr)\;
\frac{d\sigma (\gamma q \rightarrow Vq)}{dt},
\label{dsdtgp}
\end{align}
where $N_{c}=3$, $G(x_j,t)$ and $q_{f}(x_j,t)$ are the gluon and quark
distribution functions respectively, and  $W^2$ is the $\gamma p$
centre-of-mass energy squared.
The struck parton in the proton, that initiates a jet in the proton
hemisphere, carries the fraction $x_j$ of the longitudinal momentum
of the incoming proton.
For future use, we introduce the integrated quantity
\be
\left\langle \frac{d\sigma (\gamma q \rightarrow Vq)}{dt} \right\rangle
= \int _{\mri{cuts}} dx_j \,
\frac{d\sigma (\gamma p \rightarrow VX)}{dt\, dx_j}.
\label{dsdtgp1}
\ee

The partonic cross-section, characterized by the invariant
collision energy squared $\hat s = x_j W^2$ is expressed in
terms of the amplitudes $M_{\lambda\lambda'} (\hat{s},t)$,
\begin{equation}
\frac{d \sigma}{dt} = \frac{1}{16 \pi}
\left( |M_{++} (\hat s,t)|^2 +
       |M_{+0} (\hat s,t)|^2 +
       |M_{+-} (\hat s,t)|^2 \right).
\label{dsdt}
\end{equation}

In the case of an unpolarized lepton (photon) beam,
it is impossible to measure directly the helicity
amplitudes due to interference effects.
Thus, in the experiment one selects a sample of vector mesons and
measures the angular decay distribution of the
decay products \cite{ZEUS1}. The following spin density
matrix elements can be determined:

\be
r^{04} _{00} = {\langle |M_{+0}|^2 \rangle \over
        \langle |M_{++}|^2 + |M_{+0}|^2 + |M_{+-}|^2 \rangle },
\ee

\be
r^{04}_{10} = {1\over 2}
         {\langle M_{++}M^*_{+0} + M_{+-}M^*_{-0} \rangle \over
         \langle |M_{++}|^2 + |M_{+0}|^2 + |M_{+-}|^2 \rangle },
\ee

\be
r^{04} _{-11} ={\langle \repart (M_{++}M_{+-}^* ) \rangle
        \over
         \langle |M_{++}|^2 + |M_{+0}|^2 + |M_{+-}|^2 \rangle },
\ee
where $\langle ... \rangle$ denotes the integration of the
parton level quantities over partonic $x_j$ with the appropriate
cuts (c.f.\ (\ref{dsdtgp1})).

\section{The lowest order amplitude}

In the high energy limit it is convenient to
use the dipole representation for the calculation
of the scattering amplitude. In this limit the helicities of quarks,
the large light-cone components of their momenta, and
the transverse positions, are conserved in the scattering
process. The small components of light-cone momenta
are integrated out. Thus, the impact factor corresponding to
a transition of a photon with helicity $\lambda$ to a meson
with helicity $\lambda'$, due to the coupling of two gluons with
momenta $k$ and $q-k$, may be represented in the following way
\be
\Phi^{\gamma(\lambda) V(\lambda')} (\vec k,\vec q) =
\int du \int d^2 \vec r \;
[\Psi^* _{V(\lambda')} (u,\vec r)] _{\alpha\beta}\,
T (\vec k, \vec q; \vec r,u)\,
[\Psi_{\gamma(\lambda)}(u,\vec r)] _{\alpha\beta},
\label{fact01}
\ee
where $\Psi_{V}$ ($\Psi_\gamma$) denotes the vector meson (photon) wave function,
$u$ ($\bar u = 1-u$) is the quark (antiquark) longitudinal momentum
fraction, and $\vec r$ the transverse quark--antiquark separation vector.
Summation over spinor indices $\alpha$ and  $\beta$ is performed.
$T(\vec k, \vec q; \vec r,u)$ is the hard amplitude of the QCD dipole scattering, which, 
up to a constant factor, equals
\be
f^{\mri{dipole}} =
e^{i{\evec q}{\evec r}u}\left( 1 - e^{-i{\evec k}{\evec r}} \right)
\left(1 -  e^{-i({\evec q}-{\evec k}){\evec r}}   \right),
\label{dipole}
\ee
see \cite{IKSS}.
If we consider the meson photoproduction off a quark, the complete
amplitude takes the form:
\be
M_{\lambda\lambda'} (q) =
\int {d^2 \vec k }
{1 \over \vec k ^2 (\vec q - \vec k)^2}
\Phi^{\gV}(\vec k, \vec q)\, \Phi^{qq} (\vec k, \vec q),
\label{mpr}
\ee
where the quark impact factor $\Phi^{qq} (\vec k, \vec q)$
is a constant. We choose the normalization\footnote{
Only the normalization of the full amplitude
is relevant in our calculation, so we move all non-trivial
factors into the photon--meson impact factor. Caution should
therefore be exercised if these impact factors are to be used
elsewhere. 
}
\be
\Phi^{qq} (\vec k, \vec q)=1.
\label{qimpf}
\ee

In order to produce the meson, we follow \cite{IKSS} in assuming
that the quark--antiquark pair has small transverse size (as determined
by the exchanged transverse momentum which is shared approximately
equally between the two gluons \cite{FR}).
This approximation allows us to expand the relevant hadronic
matrix element about the light-cone, keeping the leading terms
in $x^2$ (where $x$ is the space-time separation of the quark and
antiquark). The relevant light-cone wavefunctions are then taken
from the QCD analyses of \cite{BBKT,BB,BBK,IKSS}.

The photon wavefunctions in the momentum representation
were obtained using the perturbative approach in
\cite{GINZ}. 
These wavefunctions
contain spinorial structures with definite chiral parity.
The chiral even components 
(corresponding to vector $\gamma^\mu$ and axial vector 
$\gamma^\mu \gamma^5$ coupling to quarks)
yield non-zero contributions in the massless limit, and the
chiral odd ones (tensor $\sigma_{\mu\nu}$) are governed by 
the quark mass $\mq$, and vanish when $\mq \to 0$. 
Thus, it is convenient to  classify the contributions to the 
impact factors and amplitudes according to their chiral parity. 
Then we have 
\be
\Phi^{\gamma(\lambda)V(\lambda')} =  
\Phi^{\gamma(\lambda)V(\lambda')} _{\mri{even}} +  
\Phi^{\gamma(\lambda)V(\lambda')} _{\mri{odd}}  
\label{phitot}
\ee
and
\be
M_{\lambda\lambda'} = 
M^{\mri{even}}_{\lambda\lambda'} +
M^{\mri{odd}}_{\lambda\lambda'},
\label{mtot}
\ee
for all photon and meson helicities $\lambda = +,-$ and 
$\lambda' = +,-,0$.
In \cite{IKSS} the available information on the vector
meson wave functions was employed to derive both
the chiral even and the chiral odd components of the amplitudes 
(\ref{mpr}).
These amplitudes, given at the leading order, 
are the starting point for our BFKL analysis.


The chiral even impact factors for $\gamma(\lambda) V(\lambda')$
computed in \cite{IKSS} are given by
\ba
\Phi^{\gamma(+)V(0)} _{\mri{even}}  = &&
-i C_V \int \frac{ d^2{\vec r}\; du\,}{4\pi}\;
\mq\, K_1 (\mq |\vec r|)\,  
{ {\vec \epsilon}^{(+)} \cdot {\vec r} \over |\vec r|}\,
f^{\mri{dipole}}\, f_V \,(1-2u)\,\phi_{||}(u),
\label{if+0} \\
%
\Phi^{\gamma(+)V(+)} _{\mri{even}} = &&
C_V \int \frac{d^2{\vec r}\;du}{4\pi}\;
\mq |\vec r|\, K_1 (\mq |\vec r|)
\,(\vec \epsilon^{(+)}\cdot\vec \epsilon^{(+)*})
f^{\mri{dipole}}
\nonumber \\
&&
\times f_V M_V
\frac{u{\bar u}}{2}\left(\int\limits_0^u \,
\frac{dv}{{\bar v}}\,\phi_{||}(v) +
\int\limits_u^1 \,\frac{dv}{v}\,
\phi_{||}(v)   \right),
\label{if++}
\\
%
\Phi^{\gamma(+)V(-)} _{\mri{even}} = &&
- C_V \int \frac{d^2{\vec r}\;du}{4\pi}\;
\mq\, K_1 (\mq |\vec r|)\,
\frac{(\vec \epsilon^{(+)}\cdot {\vec r})
(\vec \epsilon^{(-)*}\cdot {\vec r})}{|\vec r|}\,
f^{\mri{dipole}}\,
\nonumber
\\
&&
\times f_V M_V \left( {\bar u}^2\,\int\limits_0^u
\,\frac{dv}{\bar{v}}\,\phi_{||}(v) + u^2\,\int\limits_u^1
\,\frac{dv}{v}\,\phi_{||}(v) \right),
\label{if+-}
\ea
with
\be
C_V = i \alpha_s ^2 {N_c^2 -1 \over N_c^2} eQ_V,
\label{cconst}
\ee
where $\mq$ is the quark mass, $f_V$ is the meson decay
constant, and we take the asymptotic wave function
$\phi_{||}(u)  = 6\, u(1-u)$ for the light vector
mesons. The effective quark charges in the vector mesons 
are given by
$Q_\rho = 1/\sqrt{2}$,
$Q_\omega = 1/3\sqrt{2}$,
and $Q_\phi = -1/3$.

The chiral odd impact factors for $\gamma(\lambda) V(\lambda')$
with a perturbative (QED) photon wavefunction read
\ba
\Phi^{\gamma(+)V(0)}  _{\mri{odd}} = &&
- i C_V \int \frac{ d^2{\vec r}\; du\,}{4\pi}
\mq\, K_0 (\mq |\vec r|)\,(\vec\epsilon^{(+)} \cdot \vec r)\,
f^{\mri{dipole}}
\nonumber \\
&&
\times   f_V^T \, M_V \, u\bar u \,
\left(\int\limits_u^1 \,
\frac{dv}{v}\,\phi_{\perp}(v) -
\int\limits_0^u \,\frac{dv}{\bar v}\,
\phi_{\perp}(v)   \right),
%
\label{oddif+0} \\
%
\Phi^{\gamma(+)V(+)} _{\mri{odd}} = &&
C_V \int \frac{ d^2{\vec r}\; du\,}{4\pi} \;
\mq\, K_0 (\mq |\vec r|)\, 
(\vec \epsilon^{(+)}\cdot\vec \epsilon^{(+)*}) \,
f^{\mri{dipole}} \,
f_V ^T \phi_{\perp}(u),
\label{oddif++}
\\
%
\Phi^{\gamma(+)V(-)} _{\mri{odd}} = &&
- {1\over 4} C_V \int \frac{ d^2{\vec r}\; du\,}{4\pi}
\mq\, K_0 (\mq |\vec r|)\,
f^{\mri{dipole}} \;
\nonumber \\
&&
\times
 (\vec \epsilon^{(+)}\cdot\vec r)
 (\vec \epsilon^{(-)*}\cdot\vec r)\,
 f_V^T \, M_V^2 \, 6(u\bar u)^2,
\label{oddif+-}
\ea
where $\phi_{\perp}(u) = 6 u\bar u$, and the tensor
decay constant $f_V ^T$ are introduced, and we
used the asymptotic form of distribution
functions $h_{||}(u)$, $h_3(u)$ and $\phi_{\perp}(u)$
in (\ref{oddif+-})~\cite{IKSS,BBKT}.

In what follows, the complex representation of the
two-dimensional transverse vectors will be used, 
e.g.\ $r = r_x + i r_y$. This leads to the substitutions
${\vec r}{\vec \epsilon}^{(+)} = - r / \sqrt{2}$,
$d^2 \vec r = d^2 r = dr dr^* / 2i$, $|{\vec r}| \to |r|$ etc.

Thus, at the lowest order, the chiral even amplitudes for vector
meson photoproduction are given by
\begin{eqnarray}
\label{m+0ev}
M^{\mri{even}}_{+0} = i C_V
\int\,\frac{d^2 {\vec k}} {{\vec k}^2({\vec k}-{\vec q})^2}\;
\frac{d^2 r\; du\,}{4\pi} \;
\mq\, {r \over |r|}\, K_1 (\mq |r|)\,
f^{\mri{dipole}}\, \frac{f_V}{\sqrt{2}}\,(1-2u) \, \phi_{||}(u)
\end{eqnarray}
\begin{eqnarray}
\label{m++ev}
&&
M^{\mri{even}}_{+\,+} =
C_V \int\,\frac{d^2{\vec k}} {{\vec k}^2 ({\vec k}-{\vec q})^2}\;
\frac{d^2 r\;du}{4\pi}\;\,
\mq |r|\, K_1 (\mq |r|)
f^{\mri{dipole}}\, f_V M_V
\nonumber \\
&&\times \frac{u{\bar u}}{2}\left(\int\limits_0^u \,
\frac{dv}{{\bar v}}\,\phi_{||}(v) + \int\limits_u^1 \,\frac{dv}{v}
\,\phi_{||}(v)   \right),
\end{eqnarray}
\begin{eqnarray}
\label{m+-ev}
&&M^{\mri{even}}_{+\,-} =
C_V \int\,\frac{d^2{\vec k}}{{\vec k}^2
({\vec k}-{\vec q})^2}\;\frac{d^2 r\;du}{4\pi}\;\,
\mq\, \frac{ r^2}{2 |r|}\,
K_1 (\mq |r|)\,
f^{\mri{dipole}}\,
f_V M_V
\nonumber \\
&&
\times \left( {\bar u}^2\,\int\limits_0^u
\,\frac{dv}{\bar{v}}\,\phi_{||}(v) + u^2\,\int\limits_u^1
\,\frac{dv}{v}\,\phi_{||}(v) \right),
\end{eqnarray}
and the chiral odd ones
\begin{eqnarray}
\label{m+0odd}
M^{\mri{odd}}_{+0} &=& i \frac{C_V}{2}
\int\,\frac{d^2 {\vec k}} {{\vec k}^2({\vec k}-{\vec q})^2}\;
\frac{ d^2{r}\; du\,}{4\pi}
\mq\, r \, K_0 (\mq |r|)\,
f^{\mri{dipole}}
\nonumber \\
&&
\times \sqrt{2}\, {f_V^T}\, M_V u \bar{u} \,
\left(\int\limits_u^1 \,
\frac{dv}{v}\,\phi_{\perp}(v) -
\int\limits_0^u \,\frac{dv}{\bar v}\,
\phi_{\perp}(v)   \right),
\end{eqnarray}
\begin{eqnarray}
\label{m++odd}
M^{\mri{odd}}_{+\,+} &=&
C_V\, \int\,\frac{d^2{\vec k}} {{\vec k}^2 ({\vec k}-{\vec q})^2}\;
\frac{ d^2{r}\; du\,}{4\pi}
\mq\, K_0 (\mq |r|)\,
f^{\mri{dipole}} \; f_V^T \, \phi_{\perp}(u),
\end{eqnarray}
\begin{eqnarray}
\label{m+-odd}
M^{\mri{odd}}_{+\,-} &=&
{C_V \over 8} \int\,\frac{d^2{\vec k}}{{\vec k}^2 ({\vec k}-{\vec q})^2}\;
\frac{ d^2{r}\; du\,}{4\pi}\, \mq\, r^2 \, K_0 (\mq |r|)\,
f^{\mri{dipole}} \;
f_V^T \, M_V^2\,
6(u\bar u)^2.
\end{eqnarray}

Let us consider the behaviour of the amplitudes in 
the massless limit. For all chiral even amplitudes, 
one obtains non-zero results in this limit, as 
$\mq K_1 (\mq |\vec r|) \to 1/|\vec r|$ for $\mq \to 0$.
The situation is different for the chiral odd amplitudes.
The factor of $\mq K_0 (\mq |\vec r|) 
\sim -\mq \log (\mq |\vec r|)$ implies the vanishing of
all perturbative chiral odd amplitudes in the massless
limit. However, as discussed in \cite{IKSS},
chiral symmetry breaking effects can induce a sizeable
chiral odd coupling of the photon to quarks. In particular,
the characteristic
mass parameter of the chiral odd coupling was estimated
in \cite{IKSS} to be about $2\pi^2 f_\gamma / N_c \simeq 
0.46$~GeV. Here we account for chiral symmetry breaking by
keeping in mind that the quark mass parameter need not
be small. We refer to Appendix B for a more detailed discussion of
the limit $\mq \to 0$.

We should also remark that we are able to reproduce the original
formulae of \cite{GINZ} by considering the only surviving amplitudes
in the limit that the quark and anti-quark are collinear, i.e. 
$M^{\mri{odd}}_{+\,+}$ and $M^{\mri{even}}_{+\,0}$.
We also note that if we make the further constraint that the quark and anti-quark share the meson's momentum equally, we obtain the formulae of \cite{FR} for the $M_{+\,+}$ amplitude, the only surviving amplitude in this case. 
In \cite{FP}, this amplitude was shown to be in good agreement
with the ZEUS data (after including BFKL effects) allowing us to conclude 
that a sizeable chiral 
odd contribution seems to be required by the data.


\section{The exact BFKL amplitude}\label{sec4}

\subsection{Generalities}

The BFKL kernel in the leading logarithmic approximation
exhibits, in the impact parameter representation, invariance under
conformal transformations \cite{Lipatov}.
The conformal symmetry of the kernel permits the following expansion
of the amplitude in the basis of eigenfunctions $E_{n,\nu}$ \cite{Lipatov}:
\begin{align}
M_{AB} (z,q)\;=\;
{1 \over (2\pi)^2}
\sum_{n=-\infty}^{n=\infty}
\int_{-\infty}^{\infty}
d\nu\; \frac{\nu^{2}+n^{2}/4}{[\nu^{2}+(n-1)^{2}/4][\nu^{2}+(n+1)^{2}/4]}
\nonumber
\\ \times
\exp [\chi_{n}(\nu)z]\; I^{A}_{n,\nu}(q)\,  ({I^{B}_{n,\nu} (q)})^*
\hspace{2cm}
\label{bfklampl}
\end{align}
where
\begin{equation}
\chi_{n}(\nu)\;=\; 4\repart \biggl(\psi(1)-\psi(1/2+|n|/2+i\nu)\biggr)
\end{equation}
is proportional to the eigenvalues of the BFKL kernel and
\be
z = \frac{3\alpha_{s}}{2\pi}
\ln \biggl( \frac{\hat s}{\Lambda^{2}} \biggr)
\label{zdef}
\ee
($\Lambda$ is a characteristic mass scale related to
$M_V^2$  and $|t|$).
\begin{equation}
I^{A}_{n,\nu}(q)\; = \; \int \frac{d^{2}k}{(2\pi)^{2}}
\Phi^{A}(k,q)\int
d^{2}\rho_{1}\,d^{2}\rho_{2}\; E_{n,\nu}(\rho_{1},\rho_{2})
\exp (ik\cdot \rho_{1} +i(q-k) \cdot \rho_{2}),
\label{impf}
\end{equation}
and analogously for the index $B$. The eigenfunctions are given by
\begin{equation}
E_{n,\nu}(\rho_{1},\rho_{2})\;=\;
\biggl(\frac{\rho_{1}-\rho_{2}}{\rho_{1}\rho_{2}}\biggr)^{-\widetilde\mu+1/2}
\biggl(\biggl(\frac{\rho_{1}-\rho_{2}}{\rho_{1}\rho_{2}}\biggr)^{*}
\biggr)^{-\mu+1/2}
\label{E}
\end{equation}
where $\mu=n/2-i\nu$ and $\tilde{\mu}=-n/2-i\nu$.
Here $k$ and $q$ are transverse two dimensional momentum vectors, and
$\rho_{1}$ and $\rho_{2}$ are position space vectors in the
standard complex representation.
The functions $\Phi^{A} = \Phi^{\gV}$ and
$\Phi^{B} = \Phi^{qq}$ are the impact factors.

The quark impact factor in representation (\ref{impf})
was found in \cite{MMR}, generalizing
the Mueller-Tang subtraction \cite{MT} to non-zero conformal spin;
\begin{equation}
I^{qq}_{n,\nu}(q) \;=\;
-\frac{4\pi\, i^n}{|q|}\;
\biggl(\frac{|q|^{2}}{4}\biggr)^{i\nu}\,
\biggl(\frac{q^*}{q} \biggr) ^{n/2}\,
\frac{\Gamma(1/2+n/2-i\nu)} {\Gamma(1/2+n/2+i\nu)}
\label{iq}
\end{equation}
for even~$n$ and $I^{qq}_{n,\nu}=0$ for odd~$n$.

We now compute the impact factors for the 
$A = \gamma(\lambda) V(\lambda')$ transition
for all helicity states and all conformal spins $n$.
The following integrals are to be evaluated
\begin{align}
I^{A}_{n,\nu}(q)\;=\;\int \frac{d^{2}k}{(2\pi)^{2}} \,
\Phi^ {A}(k,q)
\int d^{2}\rho_{1}\, d^{2}\rho_{2}\;
E_{n,\nu}(\rho_{1},\rho_{2}) \nonumber \\
\times \exp (ik^{*}\rho_{1}/2 +ik\rho_{1}^{*}/2 +
i(q^{*}-k^{*})\rho_{2}/2 +i(q-k)\rho_{2}^{*}/2).
\end{align}
Changing variables to $\rho_{1}=R+\rho/2$, $\rho_{2}=R-\rho/2$ and
integrating over $d^{2} k$ we get
\be
I_{n,\nu}^A (q) =
\int d^2 \rho \; \Phi^A _{q} (\rho)\
\int d^2R \; E_{n,\nu}(R,\rho) e^{iq \cdot R},
\label{master}
\ee
where
$\Phi^ A _{q} (\rho)$ is the Fourier transform of the impact
factor in the momentum space $\Phi^A(\vec k, \vec q)$,
\be
\Phi^A _{q} (\rho) = \int\frac{d^2 k}{(2\pi)^2}
\Phi^A (k, q) e^{i ( k -  q /2) \cdot \rho}.
\ee
The integration over $R$ in (\ref{master}) has been done by Navelet
and Peschanski~\cite{NP}, resulting in
\be
I_{n,\nu} ^A (q) = \int d^2 \rho \; \Phi^ A_{q}(\rho)\;
\hat E^q _{n,\nu} (\rho),
\ee
where the mixed representation eigenfunction $E^q _{n,\nu} (\rho)$ is
given by
\ba
\hat E^q _{n,\nu} &=& {\frac{(-1)^n}{ 2\pi ^2}}\, b_{n,\nu} \;
\left(\frac{|q|}{8}\right)^{2i\nu}
\left( \frac{ q^*}{q} \right)^{n/2}
\Gamma(1-i\nu-n/2)\Gamma(1-i\nu+n/2)\nonumber\\
&\times&
|\rho|\, [ J_{\mu}( q^*\rho/4) \;  J_{\widetilde\mu}(q\rho^* /4)  -
(-1)^n \; J_{-\mu}( q^*\rho /4) \; J_{-\widetilde\mu}(q\rho^* /4) ]
\ea
and
\be
b_{n,\nu} = {\frac{2^{4i\nu} \pi^3}{|n/2| - i\nu}}\;
\frac{\Gamma (|n|/2-i\nu+1/2)\Gamma (|n|/2+i\nu)}
 {\Gamma (|n|/2+i\nu+1/2)\Gamma (|n|/2-i\nu)  }.
\ee

The integral over $k$ of impact factors
(\ref{if++}, \ref{if+0}, \ref{if+-}) and
(\ref{oddif++}, \ref{oddif+0}, \ref{oddif+-})
may be easily performed
since $k$ appears only in $f^{\mri{dipole}}$. Due to the
resulting $\delta$ functions $\delta(\rho)$, $\delta(r-\rho)$ and
$\delta(r+\rho)$ the integration over $r$ is trivial.
Using the fact that $\hat E^q _{n,\nu} (\rho) = 0$ for
$\rho = 0$, and
\be
\hat E^q _{n,\nu} (\rho) = (-1)^n \hat E^q _{n,\nu} (-\rho)
\label{eparity}
\ee
one arrives, for all impact factors, at complex
integrals of the type
\ba
I_{\alpha\beta}(\nu, n, q, u;a) &=&
\mq \int  d^{2}\rho\;\rho^{\alpha+1} \rho^{*\,\beta+1}\, K_a(\mq |\rho|)
e^{\frac{i\xi}{4}[q^{*}\rho + q \rho^{*} ]}  \nonumber
\\
&\times& [J_{\mu}(q^{*}\rho/4) J_{\tilde\mu}(q\rho^{*}/4) -
(-1)^n \; J_{-\mu}(q^{*}\rho/4) J_{-\tilde\mu}(q\rho^{*}/4)]
\label{ialbe}
\ea
where $\xi=2u-1$ and $K_a(x)$ is the modified Bessel
function, displaying effects of the quark mass.
The parameter $a$ equals $1$ for the chiral-even
and $0$ for the chiral-odd impact factors of \cite{IKSS}.

\subsection{Amplitudes}

The helicity amplitudes can then be expressed in the following
way
\begin{eqnarray}
&& M_{+0}^{\mri{even}}= \frac{iC_V f_V}{4\sqrt{2}|q|}
\int_0 ^1 du \;
6\, u (1-u)\, (1-2u)
\nonumber \\ 
&& \sum_{n=-\infty}^{n=+\infty}
\int_{-\infty}^{\infty}d\nu
\frac{\nu^{2}+n^{2}}{[\nu^{2}+(n-1/2)^{2}][\nu^{2}+(n+1/2)^{2}]}
\frac{\exp [\chi_{2n}(\nu)z]}{\sin (i\pi\nu)} \,
I_{0-1}(\nu,2n, q, u;1), \nonumber \\ \label{em+0ev}
\end{eqnarray}

\begin{eqnarray}
&& M_{++}^{\mri{even}}=\frac{C_V f_V M_V}{8|q|}\,
\int_0 ^1 du\;
6\, u (1-u)\,(u^2-u+1/2)
\nonumber\\
&& \times \sum_{n=-\infty}^{n=+\infty}
\int_{-\infty}^{\infty}d\nu
\frac{\nu^{2}+n^{2}}{[\nu^{2}+(n-1/2)^{2}][\nu^{2}+(n+1/2)^{2}]}
\frac{\exp [\chi_{2n}(\nu)z]}{\sin (i\pi\nu)} \,
I_{00}(\nu,2n,q, u;1), \nonumber \\ \label{em++ev}
\end{eqnarray}

\begin{eqnarray}
&& M_{+-}^{\mri{even}}=
\frac{C_V f_V M_V}{ 8|q|} \int_0 ^1 du\, 
6\, u^2\, (1-u)^2
\nonumber \\
&& \times \sum_{n=-\infty}^{n=+\infty}\int_{-\infty}^{\infty}d\nu
\frac{\nu^{2}+n^{2}}{[\nu^{2}+(n-1/2)^{2}][\nu^{2}+(n+1/2)^{2}]}
\frac{\exp [\chi_{2n}(\nu)z]}{\sin (i\pi\nu)} \,
I_{+1-1}(\nu,2n, q, u;1), \nonumber \\ \label{em+-ev}
\end{eqnarray}
where we performed integration over $v$ in 
(\ref{m+0ev}, \ref{m++ev}, \ref{m+-ev}).

Analogously, one may express the chiral odd amplitudes
\begin{eqnarray}
&& M_{+0}^{\mri{odd}}=
\frac{iC_V f_V ^T M_V}{8\sqrt{2}|q|}
\int_0 ^1 du \;
6\, u (1-u)\, (1-2u)
\nonumber \\
&&
\times \sum_{n=-\infty}^{n=+\infty}
\int_{-\infty}^{\infty}d\nu
\frac{\nu^{2}+n^{2}}{[\nu^{2}+(n-1/2)^{2}][\nu^{2}+(n+1/2)^{2}]}
\frac{\exp [\chi_{2n}(\nu)z]}{\sin (i\pi\nu)} \,
I_{{1 \over 2}\, -{1\over 2}}(\nu,2n, q, u;0),
\nonumber\\
\label{em+0odd}
\end{eqnarray}

\begin{eqnarray}
&& M_{++}^{\mri{odd}}=\frac{C_V f_V^T}{4 |q|}\,
\int_0 ^1 du\;
6\, u (1-u)\,
\nonumber\\
&& \times \sum_{n=-\infty}^{n=+\infty}
\int_{-\infty}^{\infty}d\nu
\frac{\nu^{2}+n^{2}}{[\nu^{2}+(n-1/2)^{2}][\nu^{2}+(n+1/2)^{2}]}
\frac{\exp [\chi_{2n}(\nu)z]}{\sin (i\pi\nu)} \,
I_{-{1\over 2}\, -{1\over 2}}(\nu,2n,q, u;0),
\nonumber
\\
\label{em++odd}
\end{eqnarray}

\begin{eqnarray}
&& M_{+-}^{\mri{odd}}=
\frac{C_V f_V ^T M_V^2}{ 32|q|} \int_0 ^1 du\,
6\, u^2\, (1-u)^2
\nonumber \\
&& \times \sum_{n=-\infty}^{n=+\infty}\int_{-\infty}^{\infty}d\nu
\frac{\nu^{2}+n^{2}}{[\nu^{2}+(n-1/2)^{2}][\nu^{2}+(n+1/2)^{2}]}
\frac{\exp [\chi_{2n}(\nu)z]}{\sin (i\pi\nu)} \,
I_{{3\over 2}\, -{1\over 2}}(\nu,2n, q, u;0).
\nonumber
\\
\label{em+-odd}
\end{eqnarray}

The derivation of integrals (\ref{ialbe}) for
integer difference\footnote{The condition
of integer $\alpha - \beta$ is necessary for the amplitudes
to be single valued.} $\alpha - \beta$ is given in
Appendix~A, and the result reads
\ba
I_{\alpha\beta}(\nu,n,q,u;a) &=&
\frac{\mq}{2}\int^{C^{\prime}+i\infty}_{C^{\prime}-i\infty}
\frac{d\zeta}{2\pi i}\Gamma(a/2-\zeta)\Gamma(-a/2-\zeta)\,
\tau_q ^{\zeta} \; (i\, \text{sign}\,(1-2u))^{\alpha-\beta+n}  \nonumber \\
&\times&  \left(\frac{4}{|q|}\right)^{4}
\left[\sin\pi(\alpha + \mu + \zeta)\; \right.
B(\alpha,\mu, q^* ,u,\zeta)\,
B(\beta,\widetilde\mu,q,u^* ,\zeta) \nonumber \\
&-& (-1)^n
\sin\pi(\alpha - \mu + \zeta)\;
B(\alpha,-\mu, q^* ,u,\zeta)\,
B(\beta,-\widetilde\mu,q,u^* ,\zeta)
\left. \right]
\label{finalintegral}
\ea
where we have introduced the dimensionless parameter $\tau_q = 4\mq^2/|q|^2$
and the conformal blocks
\ba
B(\alpha,\mu, q^* ,u,\zeta) =  
(-4u \bar u)^{-(\mu+2+\alpha+\zeta)/2}
\left(\frac{4}{ q^* }\right)^\alpha
2^{-\mu}\,
\frac{\Gamma(\mu+2+\alpha+\zeta)}{\Gamma(\mu+1)} \nonumber \\
\f\left(\frac{\mu+2+\alpha+\zeta}{2} \, , \,
\frac{\mu-1-\alpha-\zeta}{2}\, ; \,
\mu+1\, ; \,\frac{1}{4u \bar u}\right).
\label{blocks2}
\ea

Note, that the sums are performed over even
conformal spins $2n$ due to properties of the quark impact
factor (\ref{iq}). It turns out that odd conformal
spin components of photon-meson impact factors
(\ref{if+0}, \ref{if++}, \ref{if+-}) also vanish.
The impact factors are symmetric under the coordinate transformation
given by ($\rho \to -\rho, \quad u \to 1-u)$.
The parity behaviour (\ref{eparity}) of conformal eigenfunctions
applied to (\ref{finalintegral}) leads to the relation
\be
I_{\alpha\beta}(\nu,n,q,u;a) = (-1)^{n+\alpha-\beta}
I_{\alpha\beta}(\nu,n,q,1-u;a).
\label{parity2}
\ee
These results may be combined to give $I_{n,\nu} ^{\gV} = 0$
for odd $n$.


The obtained expressions for the impact factors are
rather complicated and the derivation is rather
involved. Therefore we performed an independent test
of the results by comparing a direct numerical
integration of (\ref{ialbe}) over $\rho$ to the analytical 
formula (\ref{finalintegral})
for some arbitrary sets of parameters, with the
quark mass tending to zero. Full agreement
was found in all cases.

\section{Conclusions and Outlook}\label{sec:conclusions}

Helicity amplitudes for diffractive vector meson 
photoproduction off a proton were investigated using 
perturbative QCD methods,  focusing on scattering
occurring at high energies and high momentum transfer, 
for which the target proton dissociates. The process is 
described in terms of hard colour singlet exchange 
between a parton from the proton and the quark-antiquark
component of the photon. 

We re-derived all the relevant impact factors
in terms of the meson light-cone wavefunction.
The helicity amplitudes were given in the two gluon
exchange approximation with arbitrary quark mass
and classified according to their chiral parity.
We suggest that the perturbative formulae with 
the constituent quark mass reasonably approximates
the effects of chiral symmetry
breaking. 

As the scattering occurs in  
Regge kinematics, the perturbative corrections are enhanced by
large logarithms of the energy, and a full resummation of the 
leading logarithmic corrections is necessary. 
Therefore, QCD corrections to the lowest order amplitudes were
also considered.
The amplitudes of the colour singlet exchange  may then be 
described in terms of the BFKL equation. Thus, we
derived the projections of the photon-meson impact
factors on the conformal eigenstates of the BFKL kernel
for an arbitrary conformal spin. 
Both the chiral even and chiral odd impact factors  are 
obtained in terms of a single complex line integral.
Using those results, the BFKL evolved helicity amplitudes 
describing the meson photoproduction off a parton 
were given in an analytical form. These are the main results
of this paper. 

Our results are correct in the leading logarithmic
approximation (LLA). Consequently, we are unable to make reliable
predictions for the absolute normalization of the cross-sections
(due to the ambiguity in the scale $\Lambda$). Moreover,
next-to-leading logarithms \cite{Fadin:1998py,Ciafaloni:1998gs}
are well known to be important
in the case of the $t=0$ BFKL equation (they have yet to
be investigated in the non-forward case). Attempts to
incorporate at least some of the next-to-leading logarithms
have been performed and indicate that the LLA is probably
quite reasonable provided one treats the leading eigenvalue of
the kernel as an effective parameter whose value in LLA is
determined by fixing $\alpha_s$ \cite{HVM,Enberg:2001ev}.
The fixing of $\alpha_s$ is
also supported by the work of \cite{Brodsky1,Brodsky2}.

It remains to confront our calculations with
the $t$-dependence of the cross-section and the spin density matrix 
elements which have been measured 
at HERA \cite{ZEUS1,Aktas:2003zi} 
for the  $\rho$, $\phi$ and $J/\psi$ mesons. 
This study will be the subject of a second paper \cite{inprep}, 
and should provide a stringent test of the BFKL approach, in addition to 
constraining the photon and meson wavefunctions.

\acknowledgments
RE wishes to thank the Theoretical Physics Group at the University of 
Manchester for their hospitality when parts of this work was carried out.
We thank Gunnar Ingelman and Lech Szymanowski for interesting discussions. 
This research was funded in part by the UK Particle Physics and
Astronomy Research Council (PPARC), by the Swedish Research Council, and
by the Polish Committee for Scientific Research (KBN) grant no.\ 5P03B~14420.

\appendix
\section{The integral for arbitrary quark mass}

It is convenient to use the variable $\xi=u-\bar u =2u-1$ rather than $u$, and to
compute the integral for $\xi$ an arbitrary complex number and
then continue analytically into the physical region $-1<\xi<1$.
We can then write (\ref{ialbe}), using 
$2i\;d^2\rho = d\rho \, d\rho^*$, as
\be
{\cal I}_\pm = \mq \; \int \frac{d\rho \, d\rho^*}{2i} \;
\rho^{\alpha+1} \rho^{*\,\beta+1} \; K_a(\mq |\rho|) \;
 e^{\frac{i}{4} (\xi^* \rho^* q+\xi\rho q^*)}\,
J_{\pm\mu}( q^*\rho/4) \;  J_{\pm\widetilde\mu}(q\rho^* /4)
\label{integral2}
\ee
so that ${\cal I}={\cal I}_+-(-1)^n \,{\cal I}_-$.

Let us start by computing the integral (\ref{integral2}) for massive quarks.
The massless case will be obtained in a suitable limit.


The Bessel functions $J$ are defined with a cut, but the products appearing here are single-valued. To make the cut structure of the integral more explicit it is convenient to represent the Bessel functions in terms of confluent hypergeometric functions, which are individually single-valued, by using the relation~\cite{GR}
\begin{equation}
J_{\sigma-1/2}(z)=\phi(\sigma,2\sigma;-2iz)\,
\left(\frac{z}{2}\right)^{\sigma-1/2}\frac{e^{iz}}{\Gamma (\sigma+1/2)}.
\end{equation}
Inserting this into Eq. (\ref{integral2}) we have
\begin{eqnarray}
{\cal I}_+&=&
\frac{64^{i\nu}\;  q^{*\,\mu} q^{\widetilde\mu}}
{\Gamma(\mu+1)\Gamma(\widetilde\mu+1)}
\; m \int \frac{d\rho \, d\rho^*}{2i}
\rho^{\alpha+1+\mu} \rho^{*\,\beta+1+\widetilde\mu}\,
\nonumber \\
&\times&
\exp \biggl[i\,\frac{\xi+1}{4}  q^* \rho+i\,\frac{\xi^*  +1}{4}q\rho^* \biggr]
\nonumber \,K_{a}(m|\rho|) \\
&\times&
\phi\left(\mu+\tfrac{1}{2},  2\mu+1 ;-\tfrac{i}{2} q^* \rho \right)\,
\phi\left(\widetilde\mu+\tfrac{1}{2},2\widetilde\mu+1;-\tfrac{i}{2} q\rho^*
\right).
\label{int_phi}
\end{eqnarray}

The Bessel function $K_a$ can be replaced with the line integral
representation\footnote{This can be seen either by the inverse Mellin
transformation or by expressing $K_a$ as a Meijer $G$ function and
using its line integral representation.}
\be
K_a(x)=\frac{1}{2}\left(\frac{2}{x}\right)^b \int^{C+i\infty}_{C-i\infty}
\frac{dz}{2\pi i} \;
\Gamma(b/2+a/2-z) \Gamma(b/2-a/2-z) \left(\frac{x^2}{4}\right)^z
\ee
for $C\le Re[b/2+a/2],\,Re[b/2-a/2]$.
Substituted into Eq.\ (\ref{int_phi}), we now get the following integral:
\ba
{\cal I}_+&=& \frac{64^{i\nu}\; q^{*\,\mu} q^{\widetilde\mu}}
{\Gamma(\mu+1)\Gamma(\widetilde\mu+1)}
\frac{m}{2} \int^{C'+i\infty}_{C'-i\infty}\frac{d\zeta}{2\pi i}
\Gamma(a/2-\zeta) \Gamma(-a/2-\zeta)\,
\biggl(\frac{m^2}{4}\biggr)^\zeta\nonumber \\
&\times&
\int \frac{d\rho \, d\rho^*}{2i}
\rho^{\alpha+1+\mu+\zeta} \rho^{*\,\beta+1+\widetilde\mu+\zeta}\,
\exp \biggl[i\,\frac{\xi+1}{4}  q^*  \rho+i\,\frac{\xi^*  +1}{4}q\rho^*
\biggr] \nonumber \\
&\times&
\phi\left(\mu+\tfrac{1}{2},  2\mu+1 ;-\tfrac{i}{2} q^*  \rho \right)\,
\phi\left(\widetilde\mu+\tfrac{1}{2},2\widetilde\mu+1;-\tfrac{i}{2} q\rho^*
\right)
\ea
where we have changed to the integration variable $\zeta=z-b/2$.

The $\rho$ and $\rho^* $ dependence of the integrand is clearly
factorized, but to disentangle the integrals we have to consider carefully
the analytic structure. It is convenient to rescale the integration
variables,
\ba
W &=& \frac{ q^*  (\xi+1)}{4i} \rho\\
Z &=& \frac{q (\xi^* +1)}{4i} \rho^* ,
\ea
giving
\ba
{\cal I}_+&=& \frac{64^{i\nu}\; q^{*\,\mu} q^{\widetilde\mu}}
{\Gamma(\mu+1)\Gamma(\widetilde\mu+1)}
\frac{m}{2} \int^{C'+i\infty}_{C'-i\infty}\frac{d\zeta}{2\pi i}
\Gamma(a/2-\zeta) \Gamma(-a/2-\zeta)\,
\biggl(\frac{m^2}{4}\biggr)^\zeta\nonumber \\
&\times&
\int \frac{dW \, dZ}{2i}
W^{\alpha+1+\mu+\zeta}\; Z^{\beta+1+\widetilde\mu+\zeta}\; e^{-(W+Z)} \nonumber\\
&\times&
\left( \frac{4i}{ q^* (\xi+1)} \right)^{2+\alpha+\mu+\zeta}
\left( \frac{4i}{q(\xi^* +1)} \right)^{2+\beta+\widetilde\mu+\zeta}
\nonumber \\
&\times&
\phi\left(\mu+\frac{1}{2},  2\mu+1 ;\frac{2W}{\xi+1}\right)\,
\phi\left(\widetilde\mu+\frac{1}{2},2\widetilde\mu+1;\frac{2Z}{\xi^* +1} \right).
\label{WZintegral}
\ea

Let us now fix $Z$ to be a constant real number. If $Z>0$ there is a
branch cut along the positive real axis from zero to infinity in the
complex $W$ plane, and if $Z<0$ the cut is along the negative real
axis. For the integral to converge $Z$ must be positive, and so to get
a non-zero contribution we have to close the contour around the cut in
the right half-plane, finding the discontinuity across the cut.
Keeping $Z$ fixed, the $W$ integral can be written as
\be
\oint_C dW(-WZ)^u e^{-W} e^{-Z} F(W)
\ee
where $u$ is a non-integer complex exponent and $F(W)$
is the non-problematic, single-valued part of the integrand. Let the contour $C$ be
the two rays $W_\pm=re^{\pm i\delta}$, where $r$ is integrated from 0
to infinity and $\delta$ is a positive real number, anticipating the
limit $\delta\to 0$. We then have
\ba
&&\oint_C dW (-WZ)^u e^{-W} e^{-Z} F(W)\nonumber\\
&&=\lim_{\delta\to 0} \int_0^\infty dr
[(-Z W_-)^u e^{-W_-}-(-Z W_+)^u e^{-W_+}] e^{-Z}F(W)\nonumber\\
&&= 2i \sin(\pi u) \; Z^u e^{-Z} \; \int_0^\infty dr\; r^u e^{-r}F(r).
\ea
That is, factorizing the two integrals, we get an extra factor
$\sin(\pi u)$ ensuring that the expression is single-valued. 
We have chosen the direction of the contour that gives the correct 
overall sign.
Applying this to (\ref{WZintegral}), with $u=\alpha + \mu +\zeta$, and using the fact that $\mu-\widetilde\mu=n$ and $\alpha-\beta$ are both integers, leads to
\ba
{\cal I}_+&=& \frac{64^{i\nu}\; q^{*\,\mu} q^{\widetilde\mu}}
{\Gamma(\mu+1)\Gamma(\widetilde\mu+1)}
\frac{m}{2} \int^{C'+i\infty}_{C'-i\infty}\frac{d\zeta}{2\pi i}
\Gamma(a/2-\zeta) \Gamma(-a/2-\zeta)\,
\biggl(\frac{m^2}{4}\biggr)^\zeta\nonumber \\
&\times&
\left( \frac{4}{ q^* (\xi+1)} \right)^{2+\alpha+\mu+\zeta}
\left( \frac{4}{q(\xi^* +1)} \right)^{2+\beta+\widetilde\mu+\zeta}
\sin\pi(\alpha + \mu +\zeta)\; (-i)^{\alpha-\beta+n}
\nonumber \\
&\times&
\int_0^\infty dW \; W^{\alpha+1+\mu+\zeta}\; e^{-W}
\phi\left(\mu+\frac{1}{2},  2\mu+1 ;\frac{2W}{\xi+1}\right)
\nonumber \\
&\times&
\int_0^\infty dZ \; Z^{\beta+1+\widetilde\mu+\zeta}\; e^{-Z}
\phi\left(\widetilde\mu+\frac{1}{2},2\widetilde\mu+1;\frac{2Z}{\xi^* +1}\right).
\label{sepWZintegral}
\ea
The last two integrals are instances of the standard integral~\cite{GR}
\begin{eqnarray}
&& \int_{0}^{\infty} du\;e^{-u} u^{b} \phi(a,2a,ku)=\Gamma(b+1) \f(a,b+1;2a;k),
\end{eqnarray}
and when using this, (\ref{sepWZintegral}) is reduced to the single line integral
\ba
{\cal I}_+&=& \frac{64^{i\nu}\; q^{*\,\mu} q^{\widetilde\mu}}
{\Gamma(\mu+1)\Gamma(\widetilde\mu+1)}
\frac{m}{2} \int^{C'+i\infty}_{C'-i\infty}\frac{d\zeta}{2\pi i}
\Gamma(a/2-\zeta) \Gamma(-a/2-\zeta)\,
\biggl(\frac{m^2}{4}\biggr)^\zeta\nonumber \\
&\times&
\left( \frac{4}{ q^* (\xi+1)} \right)^{2+\alpha+\mu+\zeta}
\left( \frac{4}{q(\xi^* +1)} \right)^{2+\beta+\widetilde\mu+\zeta}
\sin\pi(\alpha + \mu +\zeta)\; (-i)^{\alpha-\beta+n}
\nonumber \\
&\times&
\Gamma(\alpha+2+\mu+\zeta)\,
\f\left(\mu+\frac{1}{2} ,\, \alpha+2+\mu+\zeta ;\, 2\mu+1 ;\,\frac{2}{\xi+1}\right)
\nonumber \\
&\times&
\Gamma(\beta+2+\widetilde\mu+\zeta)\,
\f\left(\widetilde\mu+\frac{1}{2} ,\, \beta+2+\widetilde\mu+\zeta ;\, 2\widetilde\mu+1 ;\,\frac{2}{\xi^* +1}\right).
\label{2f1integral}
\ea

We are now ready to express ${\cal I}$ in its final form for $n$ even,
\ba
{\cal I} &=& {\cal I}_+ - (-1)^{n}\, {\cal I}_- \nonumber\\
&=&  \frac{m}{2}\int^{C^{\prime}+i\infty}_{C^{\prime}-i\infty}
\frac{d\zeta}{2\pi i}\Gamma(a/2-\zeta)\Gamma(-a/2-\zeta)\,
\tau_q^{\zeta} \;\; \left(\frac{4}{|q|}\right)^{4} (-i)^{\alpha-\beta+n} \nonumber \\
&\times& 
\left[\sin\pi(\alpha + \mu +\zeta)\; \right.
B^\prime(\alpha,\mu, q^* ,\xi,\zeta)\,
B^\prime(\beta,\widetilde\mu,q,\xi^* ,\zeta) \nonumber \\
&-& 
\sin\pi(\alpha - \mu +\zeta)\;
B^\prime(\alpha,-\mu, q^* ,\xi,\zeta)\,
B^\prime(\beta,-\widetilde\mu,q,\xi^* ,\zeta)
\left. \right],
\label{finalintegralap}
\ea
where we have introduced the dimensionless parameter
$\tau_q = 4m^2/|q|^2$ and the conformal blocks
\ba
B^\prime(\alpha,\mu, q^* ,\xi,\zeta) =
\left(\frac{1}{\xi+1}\right)^{\mu+2+\alpha+\zeta}
\left(\frac{4}{ q^* }\right)^\alpha
2^{-\mu}\,
\frac{\Gamma(\alpha+2+\mu+\zeta)}{\Gamma(\mu+1)} \nonumber \\
\f\left(\mu+\frac{1}{2}\, ,\,
\mu+2+\alpha+\zeta \,;\, 2\mu+1 \,;\,\frac{2}{\xi+1}\right).
\label{blocks1}
\ea
We can make the replacement
\begin{eqnarray}
B^\prime(\alpha,\mu, q^* ,\xi,\zeta)\rightarrow 
B^{\prime\prime}(\alpha,\mu, q^* ,\xi,\zeta),
\end{eqnarray}
where
\ba
B^{\prime\prime}(\alpha,\mu, q^* ,\xi,\zeta) &=& \left(\frac{|\xi|}{\xi}\right)^{\alpha+\mu}\;
(\xi^2-1)^{-(\mu+2+\alpha+\zeta)/2}
\left(\frac{4}{ q^* }\right)^\alpha
2^{-\mu}\,
\frac{\Gamma(\mu+2+\alpha+\zeta)}{\Gamma(\mu+1)} \nonumber \\
&& \f\left(\frac{\mu+2+\alpha+\zeta}{2} \, , \,
\frac{\mu-1-\alpha-\zeta}{2}\, ; \,
\mu+1\, ; \,\frac{1}{1-\xi^2}\right),
\label{blocks2ap}
\ea
by using the identities 9.134:1 and 9.131:1 of~\cite{GR} and assuming $n$ is even, considering that the conformal block will be multiplied with another one. This displays explicitly that the $\xi \leftrightarrow -\xi$ symmetry depends on $\alpha$ and $\beta$;
there will be an overall factor in the product of conformal blocks,
\be 
\left(\frac{|\xi|}{\xi}\right)^{\alpha+\mu}
\left(\frac{|\xi^*|}{\xi^*}\right)^{\beta+\widetilde\mu}
= (\text{sign }\xi)^{\alpha-\beta+n},
\ee 
when analytically continuing into the physical region $-1 < \xi < 1$. This factor is pulled out of the conformal blocks in equation (\ref{blocks2}), appearing in the main body of the text.

The results (\ref{finalintegralap}) and (\ref{blocks2ap}) show features
of our earlier calculations of heavy~\cite{HVM} vector meson photoproduction
for $a=0$. For $\mq \to 0$, it gives the amplitudes
for light vector meson production in the massless case~\cite{LVM}.
In Appendix \ref{lightlimit} we will show the result in the massless quark
limit, and how to obtain it from the present massive quark result.

Note that when evaluating these expressions numerically it is
necessary to carefully consider the branch cut structure of the
conformal blocks, which are not single-valued individually. It is then
useful to further transform Eqs.\ (\ref{blocks1}) or (\ref{blocks2ap})
by e.g.\ using formula 9.132:2 of~\cite{GR}, isolating the branch cuts
outside of the hypergeometric functions.

\section{The massless quark limit}\label{lightlimit}

In~\cite{LVM} the formulae corresponding to (\ref{finalintegralap}) and
(\ref{blocks2ap}) were computed in the approximation of zero quark mass.
This corresponds to using the small argument behaviour
for the case $a=1$, relevant for the calculation of
chiral even amplitudes,
\be
m K_1 (m|\rho|) \sim 1/|\rho|
\label{masscases}
\ee
Then, one obtains a simpler result
${\cal J}={\cal I}|_{m=0}$ not involving the line
integral over $\zeta$:
%
\ba
{\cal J} &=&
\left(\frac{4}{|q|}\right)^{4} \; (-i)^{\alpha-\beta+n} \;
\left[\sin\pi(\alpha + \mu - 1/2)\;
C(\alpha,\mu, q^* ,\xi)\,
C(\beta,\widetilde\mu,q,\xi^* ) \right.\nonumber \\
&-& (-1)^n
\sin\pi(\alpha - \mu - 1/2)\;
C(\alpha,-\mu, q^* ,\xi)\,
C(\beta,-\widetilde\mu,q,\xi^* )
\left. \right],
\label{masslessintegral}
\ea
where the $C(\alpha,\mu, q^* ,\xi)$ conformal blocks are given
by\footnote{The substitution $\zeta=-1/2$ is made to keep the
definitions of $\alpha$ and $\beta$ the same, because there is a
difference by a factor of $|\rho|$ between the massless and massive
integrand.}
$B^{\prime\prime}(\alpha,\mu, q^* ,\xi, \zeta=-1/2)$
\ba
C(\alpha,\mu, q^* ,\xi) = \left(\frac{|\xi|}{\xi}\right)^{\alpha+\mu}
(\xi^2-1)^{-(\mu+3/2+\alpha)/2}
\left(\frac{4 }{ q^* }\right)^\alpha
2^{-\mu}\,
\frac{\Gamma(\mu+3/2+\alpha)}{\Gamma(\mu+1)} \nonumber \\
\f\left(\frac{\mu+3/2+\alpha}{2} \, , \,
\frac{\mu-1/2-\alpha}{2}\, ; \,
\mu+1\, ; \,\frac{1}{1-\xi^2}\right)
\label{blocks3}
\ea
The calculation proceeds in a similar way to the preceding one and we
do not show the details here. It should, however, be possible to
recover the massless result from the massive one by taking the limit
$m\to 0$. This can be done by taking the leading pole approximation
for the $\zeta$ integral in the limit $\tau_q=4m^2/|q|^2\to 0$.
%
%
The product of Gamma functions in the integrand becomes
$\Gamma(1/2-\zeta) \Gamma(-1/2-\zeta)$ and there is a simple pole at
$\zeta=-1/2$ and double poles at $\zeta=1/2, 3/2,\dots$. The double
poles, however, vanish in the limit $\tau_q\to 0$, and using the
residue theorem for the pole at $\zeta=-1/2$ reproduces the massless
result (\ref{masslessintegral}).
Note that formula (\ref{masslessintegral}) is a
generalization of an integral calculated in~\cite{NP} for
$\alpha=\beta$, and we have checked that our result agrees.

The perturbative chiral-odd photon wave function corresponds to $a=0$
and is proportional to $m K_0( m|\rho|) \to -m \ln (m|\rho|)$, thus it
vanishes in the massless limit.

We stress that the above formulae (i.e. in the massless limit) are only
reliable away from the regions of $u$ close to 0 or 1.
This is because the substitution \ref{masscases} breaks down as
$\rho \to \infty$. In effect the
expansion parameter is more like $m^2/(u(1-u)|q|^2)$ and this is
not small at the endpoints. In particular, a careful treatment reveals 
that there are no endpoint singularities in the full amplitude in
the massless limit, for any individual conformal spin. In the $z\rightarrow 0$ (two-gluon) limit, the endpoint singularities shown by \cite{IKSS} in the $M_{++}^{\text{even}}$ amplitude are recovered by a non-convergent sum over all
conformal spins, analogous to that demonstrated in \cite{MMR} for 
$qq\rightarrow qq$ scattering.

\end{document}